\documentclass[aps,pre,twocolumn,superscriptaddress]{revtex4-1}
\usepackage{graphicx}% Include figure files
\usepackage{dcolumn}% Align table columns on decimal point
\usepackage{bm}% bold math
\usepackage{amsmath}
\usepackage{verbatim}
\usepackage{amssymb}
\usepackage{listings}
\usepackage{color}
\renewcommand\vec[1]{\boldsymbol{#1}}

\newcommand\Fig[1]{Fig.~\ref{#1}}
\newcommand\quot[1]{``#1''}

\newcommand\eq[1]{Eq.~(\ref{#1})}

\begin{document}

\title{Event-chain algorithm for the Heisenberg
model: Evidence for $z\simeq 1$ dynamic scaling} 

\author{Yoshihiko Nishikawa}
\email[]{nishikawa@huku.c.u-tokyo.ac.jp}
\affiliation{Department of Basic Science, the University of Tokyo\\
3-8-1 Komaba, Meguro, Tokyo 153-8902, Japan}
\author{Manon Michel}
\email{manon.michel@ens.fr}
\affiliation{Laboratoire de Physique Statistique, Ecole Normale Sup\'{e}rieure
/ PSL Research University, UPMC, Universit\'{e} Paris Diderot, CNRS, 24 rue
Lhomond, 75005 Paris, France}

\author{Werner Krauth}
\email{werner.krauth@ens.fr}
\affiliation{Laboratoire de Physique Statistique, Ecole Normale Sup\'{e}rieure
/ PSL Research University, UPMC, Universit\'{e} Paris Diderot, CNRS, 24 rue
Lhomond, 75005 Paris, France}

\author{Koji Hukushima}
\email{hukusima@phys.c.u-tokyo.ac.jp}
\affiliation{Department of Basic Science, the University of Tokyo\\
3-8-1 Komaba, Meguro, Tokyo 153-8902, Japan}
\affiliation{Center for Materials Research by Information Integration,
National Institute for Materials Science, 1-2-1, Sengen, Tsukuba,
Ibaraki, 305-0047, Japan} 

\date{\today}
 \begin{abstract}
We apply the event-chain Monte Carlo algorithm to the three-dimensional
ferromagnetic Heisenberg model. The algorithm is rejection-free and also
realizes an irreversible Markov chain that satisfies global balance. The
autocorrelation functions of the magnetic susceptibility and the energy
indicate a dynamical critical exponent $z \approx 1$ at the critical
temperature, while that of the magnetization does not measure the performance of
the algorithm. This seems to be the first report that the event-chain Monte
Carlo algorithm substantially reduces the dynamical critical exponent from
the conventional value of $z\simeq 2$. 
\end{abstract}


\maketitle

\section{Introduction}

Ever since the advent of the local Metropolis algorithm (LMC)
\cite{Metropolis_1953}, Monte Carlo simulations of systems with many degrees of
freedom have played an important role in statistical physics. Near phase
transitions, LMC is severely hampered by dynamical arrest phenomena such as
critical slowing down for second-order transitions, nucleation and coarsening at
first-order transitions, and glassy behavior in disordered systems. 
A number of specialized algorithms then allow to speed up the sampling of
configuration space, namely the Swendsen--Wang
\cite{Swendsen_1987} and the Wolff
\cite{Wolff_1989} cluster algorithms, the multicanonical method \cite{Berg_1992}
and the exchange Monte Carlo method \cite{Hukushima_1996} based on extended
ensembles. 

The above algorithms respect detailed balance, a sufficient condition for the
convergence towards the equilibrium Boltzmann distribution.
Recently, algorithms breaking detailed balance but satisfying the necessary
global-balance condition have been discussed
\cite{Suwa_2010,Turitsyn_2011,Fernandes_2011, Bernard_2009}. Among them, the
event-chain Monte Carlo (ECMC) algorithm \cite{Bernard_2009} has proven
useful in hard-sphere \cite{Bernard_2011, Isobe_2015} and more general particle
systems \cite{Peters_2012, Michel_2014}, allowing to equilibrate larger
systems than previously possible \cite{Kapfer_2015,Isobe_2015}. It has also been
applied to continuous spin systems \cite{Michel_2015}. ECMC uses a factorized
Metropolis filter \cite{Michel_2014} and relies on an additional \quot{lifting}
variable to augment configuration space \cite{Diaconis_2000}. It is
rejection-free and realizes an irreversible Markov chain.
So far, however, the speedup realized by ECMC with respect to LMC has always
represented a constant factor in the thermodynamic limit, although larger gains
are theoretically possible \cite{Diaconis_2000,Chen_1999}. 

In this paper, we apply ECMC  to the three-dimensional ferromagnetic Heisenberg
model, defined by the energy
\begin{equation}
\label{eq:Hamiltonian}
 E\left( \left\{ \vec{S}_i\right\}\right) = - J\sum_{\langle i, j\rangle}
\vec{S_i} \cdot \vec{S}_j,
\end{equation}
where $J$ is the unit of the energy, $\vec{S}_i$ is a three-component unit vector and the
sum runs over all neighboring pairs of the $N = L^3$ sites of a simple cubic
lattice of linear extension $L$.  In our simulations, we consider the critical
inverse temperature $\beta_\mathrm{c}  = J / T_\mathrm{c} = 0.6930$
\cite{StaticFerroCFL}. To describe the dynamics of
the system, we compute the autocorrelation functions of the energy, the system
magnetization $\vec{M}=  \sum_k \vec{S}_k$
and the magnetic susceptibility
\begin{equation}
 \chi = \frac{| \vec{M} | ^2}{N}.
\end{equation}
The energy and the susceptibility are both invariant under global
rotations of the spins $ \vec S_k$ around a common axis, whereas the
magnetization follows the rotation. We will argue that the energy and the
susceptibility are slow variables, that is that their slowest time constant
describes the correlation (mixing) time of the underlying Markov chain.
Under this hypothesis, we will present evidence that the ECMC for the
three-dimensional Heisenberg model reduces the dynamical critical exponent from
the LMC value of $z\simeq 2$ to $z\simeq 1$. This considerable reduction of
mixing times with respect to the LMC  may well be optimal within the lifting
approach \cite{Chen_1999}. The observed reduction is all the more surprising as
in the closely related $XY$ model \cite{Michel_2015}, where the spins are
two-dimensional unit vectors, the ECMC realizes speedups by two orders of
magnitude with respect to LMC, but does not seem to lower the dynamical critical
exponent.

\section{ECMC algorithm for the Heisenberg model}

\begin{figure*}[t]
\includegraphics[width=0.32\textwidth]{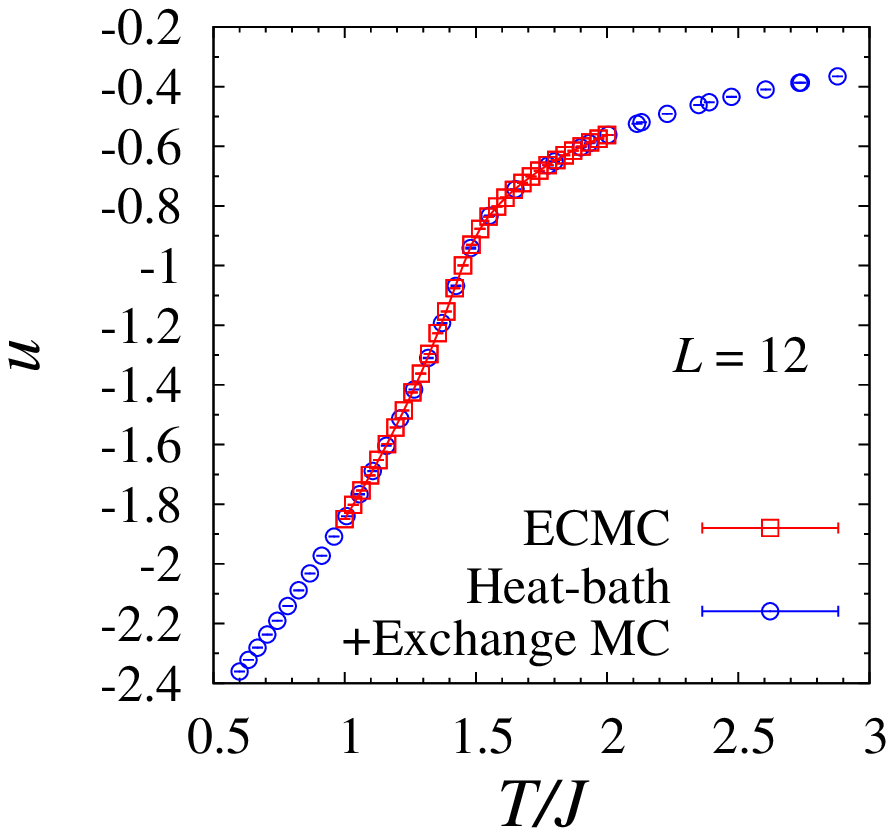}
\includegraphics[width=0.32\textwidth]{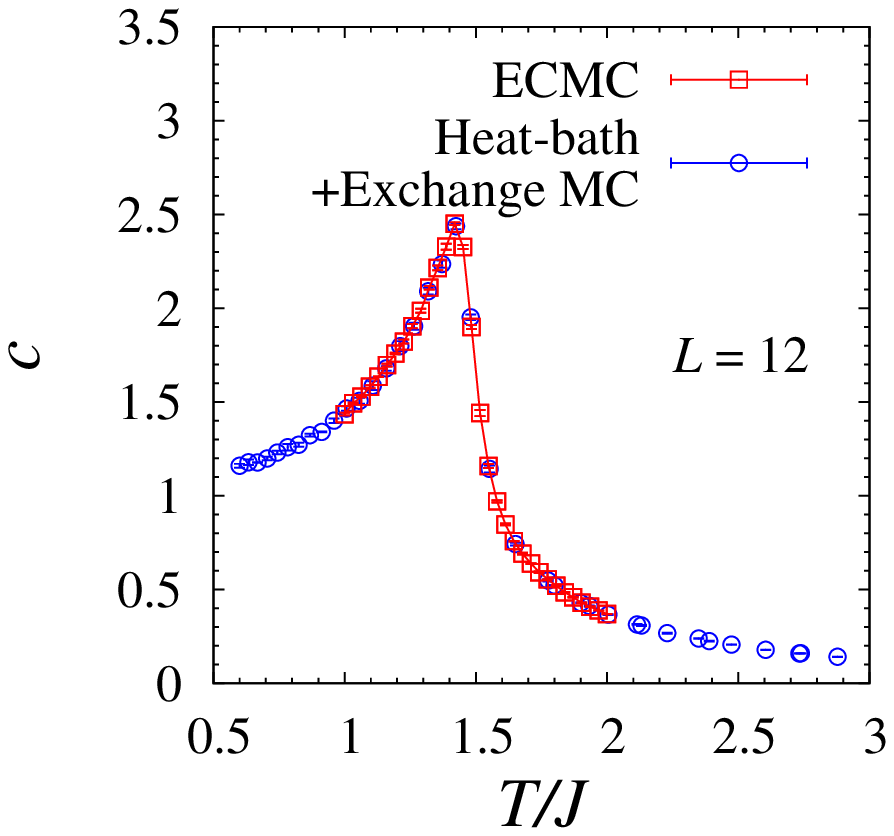}
\includegraphics[width=0.32\textwidth]{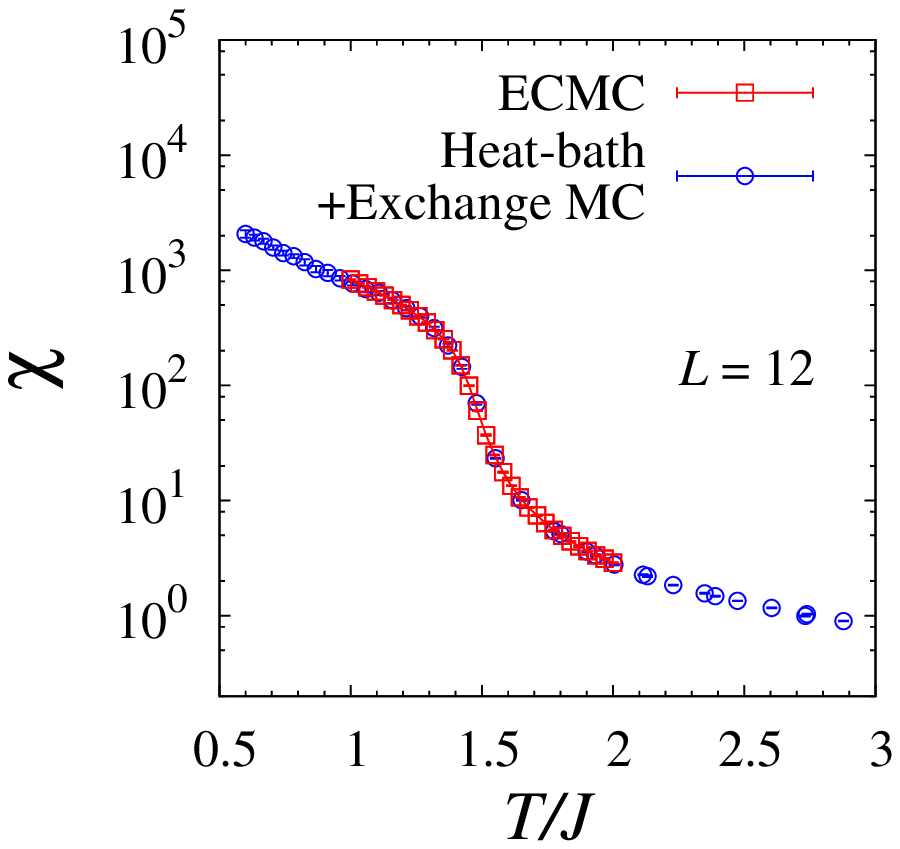}
\caption{(Color online) 
Temperature dependence of the energy density $e = E/N$, the specific heat $c$
and the magnetic susceptibility $\chi$ of the three-dimensional Heisenberg
 model with $L = 12$. A chain length $\ell = N\pi/10$ is used.}
\label{fig:static}
\end{figure*} 
Applied to the Heisenberg model, the ECMC augments the physical space of spin
configurations by a lifting variable $(k, \vec v)$ which specifies the
considered infinitesimal counterclockwise rotation of spin $k$  about the axis
$\vec v$. By virtue of the factorized Metropolis filter, this physical move can
only be rejected by a single neighboring spin, $l$,  and the lifting variable
will then be moved as $(k, \vec v) \to (l, \vec v)$, keeping the sense of
rotation, but passing it on to the spin responsible for the rejection. In the
augmented space, the rejections are thus supplanted by events, namely the
lifting moves for arrested  physical states. The ECMC, for a given axis
$\vec v$, breaks detailed balance, yet satisfies global balance, as the
probability flow into each lifted configuration equals the flow out of it, to
first order in the time increment $\mathrm{d}t$. For the $XY$ model of planar rotators,
$\vec  v$ is uniquely defined as the axis perpendicular to the sense of
rotation.
For this reason, the ECMC around this axis is irreducible, and the chain length
$\ell$ in this model is best taken equal to the simulation
time \cite{Michel_2015}. For the Heisenberg model, spin rotations must be about
at least two axes, in order to reach the entire configuration space. The
resampling of the rotation axis is performed after the cumulative rotation
angles about the previous axis reaches the chain length $\ell$. All
configurations of the chain sample the equilibrium distribution and any uniform
subset of them yield valid observable averages. Observables may be integrated 
during the continuous evolution or {\em e.g.} retrieved at regular intervals
independent of the lifting events.

For a fixed rotation axis $\vec v$, the ECMC algorithm for the Heisenberg model
reduces to the one of the $XY$ model: With $(\phi_{\vec v, k}, \theta_{\vec v,
k})$ the spherical coordinates of a spin $k$ in a system where the $z$-axis is 
aligned with $\vec v$, 
the pair energy $E_{kl}$ between spins $k$ and $l$ is
\begin{align}
E_{kl} &= -J'\cos( \phi_{\vec v, k} - \phi_{\vec v, l}) + K
\label{HXY_E}
\intertext{with}
J' &= J\sin \theta_{\vec v, k}\sin \theta_{\vec v, l}, \notag\\
K &= - J\cos \theta_{\vec v, k} \cos \theta_{\vec v, l}. \notag
\end{align}
Both $J'$ and $K$ depend only on the polar angles $\theta_{\vec v}$ and
remain unchanged along the event chain. The azimuthal-angle dependence in
\eq{HXY_E} is $\propto \cos (\phi_{\vec v, k} - \phi_{\vec v, l})$, as in the
$XY$ model. 

The azimuthal angle $\phi_{\vec v, k}$ increases for each  ECMC chain from its
initial value $\phi_0$ until one of its neighbors, $l$,  triggers a lifting $(k,
\vec v) \to (l, \vec v)$ at $\phi_{\vec v, k} = \phi_{l,\text{event}}$.
The latter is sampled with a single random number in the event-driven approach
\cite{Peters_2012,Michel_2014}. Precisely, $\phi_{l, \text{event}}$ is given by
the sampling of the positive pair energy increase:
\begin{multline}
\Delta E_l = - \left[\log\ \text{ran}(0, 1)\right]/ \beta = \\
-J'\int^{\phi_{l, \text{event}}}_{\phi_0}
             \max\left(0, \frac{\mathrm{d}\cos(\phi_{\vec v, k} - 
             \phi_{\vec v, l})}{\mathrm{d}\phi_{\vec v, k} } \right) \mathrm{d}\phi_{\vec v, k},
\label{Sampling_phi}
\end{multline}
where $\text{ran}(0, 1)$ is a uniform random number between $0$
and $1$.
To solve \eq{Sampling_phi} for $\phi_{l, \text{event}}$,  one first slices off 
any full rotations (these $n$ rotations by $2 \pi$ yield 
an energy increase of $ 2 n J'$),
leaving a value $\Delta E_l^f$,
\begin{equation}
E^*_{\text{init}} + \Delta E_l^f  = -J'  \cos( \phi_{l, \text{event}} -
\phi_{\vec v, l} - 2n\pi),
\label{inversion_phi}
\end{equation}
where
\begin{equation*}
E^*_{\text{init}} = 
\begin{cases}
E_{kl} &\text{if initial pair energy change $> 0$} \\ 
-J'    &\text{otherwise}.
\end{cases}
\end{equation*}
The true lifting event corresponds to the earliest of the independent event
times sampled for all the neighbors of the spin $k$. In ECMC, Monte Carlo
time is continuous and proportional to the total displacement of the spins.  
We have checked the correctness of the
ECMC, and obtained perfect agreement for the mean energy, the specific heat and
the susceptibility with the heat-bath algorithm \cite{Miyatake_1986,Olive_1986}
modified with the exchange Monte Carlo method (or ``parallel tempering'')
\cite{Hukushima_1996} (see Fig.~\ref{fig:static}).

\section{Dynamical scaling exponent}

\begin{figure*}[htb]
\begin{center}
\includegraphics[width=1.0 \textwidth]{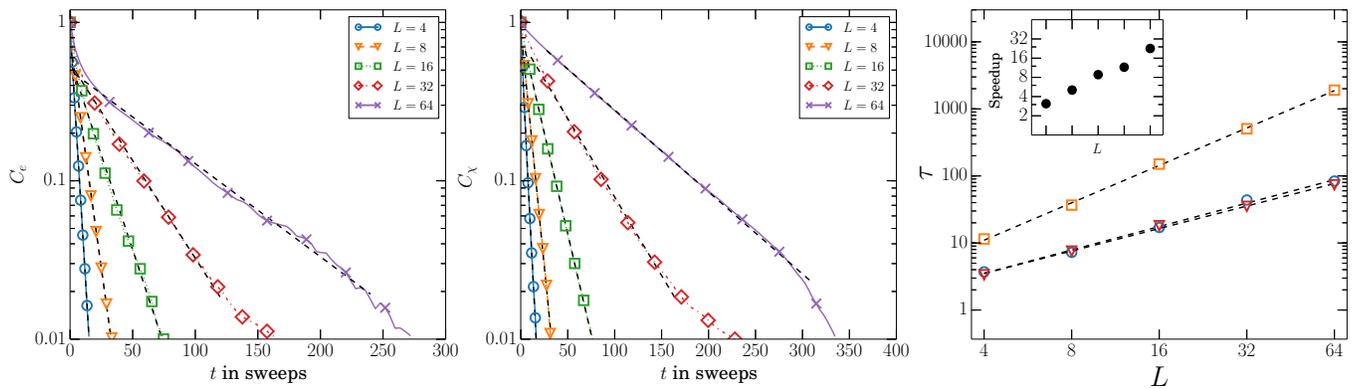}
\caption{Autocorrelation functions and time constants of ECMC for the
three-dimensional Heisenberg model at its critical point $\beta = 0.693$.
  \emph{Left:} Energy density autocorrelation function $C_e$ for for
  system sizes $4^3$, $8^3$, \dots, $64^3$.  \emph{Center:}
  Susceptibility autocorrelation function $C_\chi$ for ECMC for the
  three-dimensional Heisenberg system sizes $4^3$, $8^3$, \ldots,
  $64^3$.  \emph{Right:} Scaling of the autocorrelation time
  $\tau_\chi$ (resp. $\tau_e$) of the susceptibility $\chi$
  (resp. energy density $e$) with system size $L$ for ECMC (\emph{blue
     circles}) (resp. (\emph{red triangles})) and of the autocorrelation time of
the susceptibility for LMC (\emph{yellow squares}). Error bars are smaller than
the markers size. \emph{Right Inset: }
  Speedup for the susceptibility $\chi$ in comparison to LMC for
  system sizes $4^3$, $8^3$, \dots, $64^3$.}
\label{f:EC_Aco}
\end{center}
\end{figure*}

 \begin{figure}[ht]
\includegraphics[width=1.0\columnwidth]{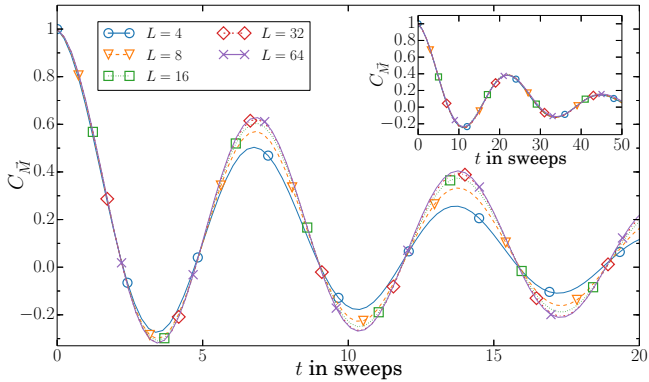}
  \caption{(Color online) Autocorrelation function of magnetization $C_M(t)$ at
the critical temperature for various system sizes. The inset shows the spin
autocorrelation function of a trivial algorithm that only performs global
rotations in spin space along the two axes.}
 \label{fig:autocorrelation}
 \end{figure}

At the critical temperature $T_\mathrm{c}$, the correlation length $\xi$
of a model undergoing a second-order phase transition equals the system size $L$
and the autocorrelation time of slow variables $\tau$ diverges as $\tau \sim L^z$, 
where $z$ is the dynamical critical exponent. We measure time in sweeps: One
ECMC sweep corresponds, in average, to $N$ lifting events and one LMC sweep
to $N$ attempted moves. Time autocorrelation functions are defined by

\begin{eqnarray}
C_O(t) &=& \frac{\left\langle O(t' + t) O(t ')\right\rangle -
\left\langle O(t') \right\rangle ^2}
{\left\langle O^2(t') \right\rangle
- \left\langle O(t') \right\rangle ^2},
\label{eq:C_m}
\end{eqnarray}
where the brackets $\langle \cdots \rangle$ indicate the thermal average
and $t'$ is set sufficiently large for equilibration. 
The dynamical critical
exponent of LMC for the three-dimensional Heisenberg
model was estimated from the autocorrelation function of the magnetization $\vec
M$ as $z = 1.96(6)$\cite{Peczak_1993}. The over-relaxation algorithm
\cite{Creutz_1987,Brown_1987} seems to give $z\simeq 1.10$ \cite{Peczak_1993} 
which was obtained from the autocorrelation function of the
magnetization,  and the Wolff algorithm is believed to yield a value close to
zero: $z \gtrsim 0$, a value obtained from the susceptibility autocorrelation
function \cite{Holm_1993}.

To evaluate the correlation time and the dynamical critical exponent for the
ECMC, one must pay attention to the irreversible nature of the underlying Markov
chain. During one event chain, spins all rotate in the same sense, and the
system undergoes global rotations with taking into account the thermal
fluctuation. This results in fast oscillations
of the magnetization $\vec M$ and a quick decay of its autocorrelation
function that is insensitive to the system size (see
Fig.~\ref{fig:autocorrelation}), and even to the temperature. However, this
effect is also  visible for a trivial algorithm, which simply performs
global rotations (see the inset of Fig.~\ref{fig:autocorrelation}). The trivial
algorithm satisfies global balance, but its correlation time is infinite, as it
does not relax the energy. A similar effect appears in the ECMC for particle
systems \cite{Bernard_2009}, that likewise is not characterized by the mean
net displacement of particles. To characterize the speed of the ECMC, we
consider the energy density and the susceptibility that we conjecture to be
slow variables at the critical temperature. Both $\chi$ and $e$ are insensitive
to global rotations and do not oscillate.

As shown in \Fig{f:EC_Aco}, the autocorrelation functions both of the
energy density and of the susceptibility are well approximated as a
single exponential decay 
\begin{equation}
C_\chi(t) = \exp(-t/\tau)
\end{equation}
on essentially the same timescales. Furthermore,  the finite-size behavior of
the autocorrelation times indicates $z\simeq 1$ dynamical scaling. This $z$
value is significantly less than for the LMC and very similar to the one
obtained for over-relaxation methods, although the $z \simeq 0$ value of the
cluster algorithm is not reached.

\section{Discussion and summary}

The earliest application of lifting \cite{Diaconis_2000},
the motion of a particle on a one-dimensional $N$-site lattice with
periodic boundary conditions, already featured the decrease of  the dynamical
scaling exponent from $z=2$ to $z=1$ (the reduction of the mixing time from
$\propto N^2$ to $\propto N$). 
To reach such reductions, the Markov chain must be irreversible.
It was pointed out that the \quot{square-root} decrease of the
critical exponent was the optimal improvement~\cite{Chen_1999}. The
concepts of factorized Metropolis filters and of infinitesimal moves 
bring irreversible lifting algorithms
to general $N$-body systems, although only finite speed-ups were realized so far
in the $N \to \infty $ limit. The three-dimensional Heisenberg model
seems to be a first such ECMC application with a lowered
critical dynamical exponent. Our observation relies on the hypotheses
that the energy and the susceptibility are indeed \quot{slow} variables, and
that the observed decay of the autocorrelation function continues for larger
times. However, in \Fig{f:EC_Aco}, a crossover from $z=1$ back to $z=2$
as it was observed in the $XY$-model after $\sim 5$ sweeps~\cite{Michel_2015}
appears unlikely to arise after hundreds of sweeps. The dynamical critical
exponent $z \approx 1$ represents a maximal improvement with respect to the $z
\approx 2$ of LMC, supposing again that the theorems of ref.~\cite{Chen_1999}
apply to infinitesimal Markov chains. 

In summary, we have successfully applied ECMC to the Heisenberg model in three
dimensions. ECMC shows considerable promise for spin models, 
and  the numerical data presented in this paper allow us to formulate
the exciting conjecture that the dynamical critical exponent for the Heisenberg
model is $z \simeq 1$. 
ECMC  is also applicable to frustrated magnets and spin glasses, which involve
antiferromagnetic interactions and/or quenched disorder.  Our preliminary study
indicates that the ECMC algorithm is also useful for a Heisenberg spin glass
model.  ECMC can be easily combined with other algorithms such as the exchange
Monte Carlo method and the over-relaxation algorithm in the usual manner.  This
may allow to investigate the three-dimensional Heisenberg spin glass model in
the low-temperature region. Large-scale simulations in this direction are
currently in progress. It would be very interesting to understand why ECMC is so
much more successful in the Heisenberg model than both in hard and soft disks
and in the $XY$ model. 

\section*{acknowledgment}

YN and KH thank S.~Hoshino and M.~J.~Miyama for useful
discussions, and J.~Takahashi and Y.~Sakai for carefully reading the
manuscript.  This research was supported by Grants-in-Aid for Scientific
Research from the {JSPS}, Japan (Nos. 25120010  and 25610102), and
JSPS Core-to-Core program ``Nonequilibrium dynamics of soft matter and
information.'' 
This work was granted access to the HPC resources of
MesoPSL financed by the Region Ile de France and the project Equip@Meso
(reference ANR-10-EQPX-29-01) of the programme Investissements d'Avenir
supervised by the Agence Nationale pour la Recherche.

\end{document}